\documentclass[AMA,STIX1COL]{WileyNJD-v2}

\usepackage[utf8]{inputenc}
\usepackage{nomencl}
\usepackage{amsfonts}
\usepackage[bbgreekl]{mathbbol}
\usepackage{etoolbox}
\usepackage{moreverb}

\usepackage{graphicx}
\usepackage{epstopdf}

\usepackage{subfig}
\usepackage{amssymb}
\usepackage{tikz,tikzscale}
\usepackage{pgfplots}
\pgfplotsset{compat=1.12}
\usepackage{pgfplotstable} 
\usepackage{pifont}
\usepgfplotslibrary{patchplots}

\newcommand{\tns}[2][0]{
\ifnum#1=0
 #2  
 \else
   \ifnum#1=1
     \boldsymbol{#2} 
   \else
     \ifnum#1=2
       \boldsymbol{#2}  
     \else
       \ifnum#1=3
         \boldsymbol{\mathcal{#2}}  
       \else
         \ifnum#1=4
           \mathbb{#2}  
         \else
           Holzapfel will kill you!
         \fi
       \fi
     \fi
   \fi
 \fi
 }
 
\newcommand{\mtns}[2][0]{
\ifnum#1=0
 #2  
 \else
   \ifnum#1=1
     \boldsymbol{#2} 
   \else
     \ifnum#1=2
       \boldsymbol{#2}  
     \else
       \ifnum#1=3
         \boldsymbol{\mathcal{#2}}  
       \else
         \ifnum#1=4
           \mathbb{#2}  
         \else
           Holzapfel will kill you!
         \fi
       \fi
     \fi
   \fi
 \fi
 }
 
\usepackage{mathtools}
\usepackage[acronyms,shortcuts]{glossaries} 
\usepackage{listings}
\usepackage{placeins}
\usepackage{booktabs}
\usepackage{xcolor, colortbl}
\usepackage{siunitx}
\usepackage{soul}
\usepackage{tcolorbox}
\usepackage{xspace}
\usepackage{longtable}

\definecolor{myblue}{RGB}{0,100,200}
\definecolor{myred}{RGB}{204,102,0}

\newtcolorbox{mybasecolorbox}[1][]{%
  colback=gray!25, colframe=gray!25,
  coltitle=black, fonttitle=\bfseries,
  sharp corners,
  width=(\linewidth-30pt),
  title=#1}

\newenvironment{mytitlebox}[1][]{%
  \centering
  \mybasecolorbox[#1]
  \itshape
}{%
  \endmybasecolorbox
}
\newenvironment{mission}{%
  \mytitlebox[Mission statement]
}{%
  \endmytitlebox
}

\DeclarePairedDelimiterX{\infdivx}[2]{}{}{%
  #1\;\big\|\;#2%
}

\newacronym{ac:uq}{UQ}{uncertainty quantification}

\newacronym{ac:hpc}{HPC}{high-performance computing}

\newacronym{ac:cae}{CAE}{computer-aided engineering}

\newacronym{ac:lhs}{LHS}{Latin hypercube sampling}

\newacronym{ac:mc}{MC}{Monte Carlo}

\newacronym{ac:qoi}{QoI}{quantity of interest}

\newacronym{ac:api}{API}{application programming interface}

\newacronym{ac:aws}{AWS}{Amazon Web Service}

\newacronym{ac:ectwo}{EC2}{Elastic Compute Cloud}

\newacronym{ac:ecs}{ECS}{Elastic Container Service}

\newcommand{\QUEENS}{\emph{QUEENS}}

\newcommand{\eg}{e.\,g., \xspace}
\newcommand{\ie}{i.\,e.,\xspace}

\newcommand{\repo}{\url{https://github.com/queens-py/queens}}
\DeclareSymbolFontAlphabet{\mathbb}{AMSb}
\DeclareSymbolFontAlphabet{\mathbbl}{bbold}

\articletype{Article Type}%

\received{<day> <Month>, <year>}
\revised{<day> <Month>, <year>}
\accepted{<day> <Month>, <year>}

\definecolor{codegreen}{rgb}{0,0.6,0}
\definecolor{codegray}{rgb}{0.5,0.5,0.5}
\definecolor{codepurple}{rgb}{0.58,0,0.82}
\definecolor{backcolour}{rgb}{0.95,0.95,0.92}
\definecolor{lightgray}{rgb}{0.95,0.95,0.95}

\lstdefinestyle{mystyle}{
    backgroundcolor=\color{backcolour},   
    commentstyle=\color{codegreen},
    numberstyle=\tiny\color{codegray},
    stringstyle=\color{codepurple},
    basicstyle=\footnotesize,
    breakatwhitespace=false,         
    breaklines=true,                 
    captionpos=b,                    
    keepspaces=true,                 
    numbers=left,                    
    numbersep=5pt,                  
    showspaces=false,                
    showstringspaces=false,
    showtabs=false,                  
    tabsize=2,
    aboveskip=0pt,
    belowskip=0pt,
    framerule=0pt,
    xleftmargin=1pt,
    xrightmargin=1pt,
    breakindent=0pt,
    resetmargins=true
}
 
\hypersetup{
	colorlinks,
	linkcolor=myred,
	citecolor=myblue,
	urlcolor=myblue
}

\begin{document}
\title{\texorpdfstring{QUEENS: An Open-Source Python Framework for Solver-Independent Analyses of Large-Scale Computational Models\\ \large{-- From Parameter Studies and Identification, Sensitivity Analysis, Surrogates, Optimization, (Bayesian) Forward and Backward Uncertainty Quantification to Digital Twinning --}}{QUEENS: An Open-Source Python Framework for Solver-Independent Analyses of Large-Scale Computational Models \large{-- From Parameter Studies and Identification, Sensitivity Analysis, Surrogates, Optimization, (Bayesian) Forward and Backward Uncertainty Quantification to Digital Twinning --}}}

\author[1,3]{J. Biehler*}
\author[1]{J. Nitzler*}
\author[1,2]{S. Brandstaeter*}
\author[1]{M. Dinkel}
\author[3]{V. Gravemeier}
\author[1]{L. J. Haeusel}
\author[1]{G. Robalo Rei}
\author[1]{H. Willmann}
\author[1]{B. Wirthl}
\author[1]{W. A. Wall}

\authormark{The \QUEENS\  developers}

\address[1]{\orgdiv{Institute for Computational Mechanics}, \orgname{Technical University of Munich}, \orgaddress{\street{Boltzmannstr. 15}, \city{Garching}, \country{Germany}}}

\address[2]{\orgdiv{Institute for Mathematics and Computer-Based Simulation}, \orgname{University of the Bundeswehr Munich}, \orgaddress{\street{Werner-Heisenberg-Weg 39}, \city{Neubiberg}, \country{Germany}}}

\address[3]{\orgdiv{AdCo Engineering$^\text{GW}$ GmbH}, \orgaddress{\street{Oskar-Messter-Str. 33}, \city{85737 Ismaning}, \country{Germany}}}

\corres{B. Wirthl\\ \email{barbara.wirthl@tum.de}}

\abstract[Abstract]{%
A growing challenge in research and industrial engineering applications is the need for repeated, systematic analysis of large-scale computational models, for example, patient-specific digital twins of diseased human organs:
The analysis requires efficient implementation, data, resource management, and parallelization, possibly on distributed systems.
To tackle these challenges and save many researchers from annoying, time-consuming tasks, we present \QUEENS\ (Quantification of Uncertain Effects in Engineering Systems), an open-source Python framework for composing and managing simulation analyses with arbitrary (physics-based) solvers on distributed computing infrastructures. 
Besides simulation management capabilities, \QUEENS\ offers a comprehensive collection of efficiently implemented state-of-the-art algorithms ranging from routines for convergence studies and common optimization algorithms to more advanced sampling algorithms for uncertainty quantification and Bayesian inverse analysis. Additionally, we provide our latest cutting-edge research in multi-fidelity uncertainty quantification, efficient multi-fidelity Bayesian inverse analysis, and probabilistic machine learning. \QUEENS\ adopts a Bayesian, probabilistic mindset but equally supports standard deterministic analysis without requiring prior knowledge of probability theory.
The modular architecture allows rapid switching between common types of analyses and facilitates building sophisticated hierarchical algorithms.
Encouraging natural incremental steps and scaling towards complexity allows researchers to consider the big picture while building towards it through smaller, manageable steps. 
We aim to create a large scientific community that seeks to collect, extend, and develop methods at the intersection of physics-based simulations, data-driven machine-learning techniques, and sophisticated simulation analytics.
The open-source repository is available at \repo.
}

\maketitle

\section{Introduction}
\label{sec: intro}
Computational models have become indispensable in advancing scientific and engineering disciplines. These models encompass many facets, such as hardware dependencies, physical model assumptions, algorithmic complexity, numerical configurations, and various (uncertain) parameterizations. 

Researchers often seek to scrutinize and optimize various aspects of computational models and their setups. This iterative process can quickly become cumbersome and often includes formidable challenges and infrastructure constraints. Moreover, many sources of uncertainty exist, including model errors, incomplete information, and inherent variability in the studied system. Understanding and quantifying these uncertainties become vital prerequisites for informed decision-making. 

Consequently, a wide range of libraries for different algorithms and types of model analysis, such as PyDOE \cite{pydoe}, SALib \cite{Iwanaga2022}, Chaospy \cite{feinberg_chaospy_2015}, UQLab \cite{marelli2014-UQLabFrameworkUncertainty}, UQPy\cite{olivier2020-UQpyGeneralPurpose}, CUQIpy \cite{riis2024cuqipy, alghamdi2024cuqipy}, ELFI \cite{Lintusaari2018}, PyMC \cite{pymc2023}, and particles \cite{chopin_introduction_2020} have emerged. 
While the significance of comprehensive model analysis is recognized, there is still a lack of suitable and robust interfaces to employ the mentioned libraries in conjunction with complex HPC simulation models and workflows stemming from decades of research. 

To overcome these challenges, we introduce the novel framework \QUEENS\ (Quantification of Uncertain Effects in Engineering Systems): An open-source Python framework for coordinating and composing complex analysis workflows for arbitrary simulation software and multi-query analysis with strong support for probabilistic methods.  The software has two goals: 
First, we want to provide a framework that facilitates a fast setup for simulation analysis that can easily scale from the personal laptop to distributed HPC workflows while taking care of all simulation and data management.
Second, \QUEENS\ is a computational platform to maintain, develop, and share algorithmic advances in optimization, uncertainty quantification, sensitivity analysis, (Bayesian) inverse analysis (backward uncertainty quantification), and simulation analytics. Most users of \QUEENS\ might use the provided abstraction levels to design their simulation analysis pipelines with their own or commercial third-party simulation frameworks, such as finite element software, particle methods, or any other numerical discretization approach. In this context, we use the terms finite element simulations or physics-based simulations as umbrella terms that should include all approaches for discretized PDE solvers. For this purpose, we already offer many state-of-the-art algorithms and our own developments  (see Section \ref{sec:overview_analysis} for an overview). Additionally, we also want to encourage active method development.
A core principle of \QUEENS\ is encouraging natural incremental scientific steps towards more complexity, which allows researchers to develop complex algorithms and workflows robustly. The study of arbitrary simulation codes requires robust infrastructure and interfaces, powerful packages like \QUEENS\ or similar state-of-the-art frameworks, such as UM-Bridge \cite{seelinger2023-UMBridgeUncertaintyQuantification}, Dakota  \cite{dakota}, OpenTURNS \cite{baudin2017-OpenTURNSIndustrialSoftware}, and EasyVVUQ \cite{Richardson2020_easyvvuq} have proven instrumental in meeting these needs. \QUEENS, along with these packages, capitalizes on a common fundamental characteristic shared by various model analyses: the repetitive evaluation of the system of interest, often in a batch sequential setting. In this context, the workflow alternates between sequential steps that are interdependent and parallel steps that can be executed simultaneously. Specifically, the parallel workload typically involves independent evaluations of the physical model, lending itself to an embarrassingly parallel algorithm that optimizes computational resources. We define such a setting in a broad sense as \emph{multi-query tasks}, \ie the efficient evaluation of model output quantities of interest, enabling the simultaneous exploration and characterization of the system under investigation.

\begin{figure}[htbp]
    \centering
    \includegraphics[width=\textwidth]{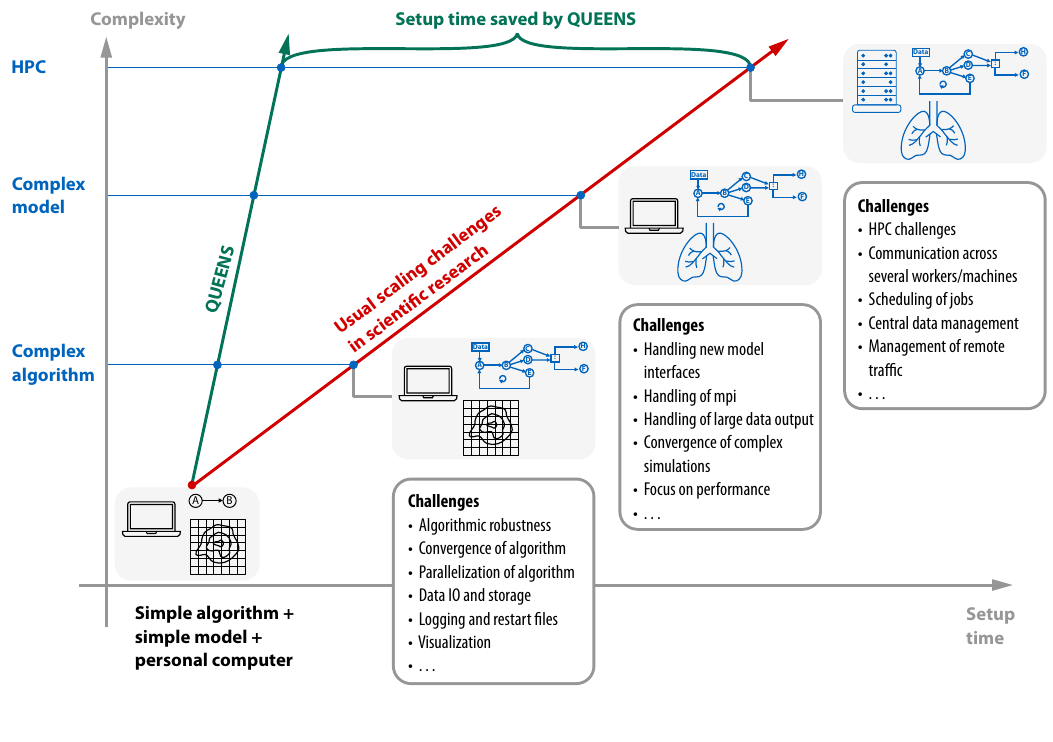}
    \caption{Usual challenges encountered when scaling scientific research workflows from \emph{personal computer - simple algorithm - simple simulation model} to \emph{distributed computing - complex algorithm - large scale, complex simulation models}. The complexity scaling is exemplified for the order: complex algorithm - complex model - HPC, but should be understood interchangeably.}
    \label{fig: scaling_challenges}
\end{figure}

When setting up analysis pipelines for existing simulation frameworks, such as finite element codes, researchers usually start their work on a personal workstation or laptop using a simplified version of the simulation model and a first draft of an algorithmic implementation for the model study. Implementing this first simple workflow is usually straightforward, but specialized to the problem at hand, needing more flexibility and abstraction for building up to more challenging workflows. Nevertheless, the goal is primarily to scale an existing setup to an HPC system using the full high-fidelity physical simulation model. Elevating the research setup to the subsequent complexity level will require many manual steps, such as updating scripts, changing interfaces, and restructuring code and algorithms to facilitate the more complex requirements. Figure \ref{fig: scaling_challenges} represents some examples of time-consuming pitfalls and challenges when setting up typical research stages. Schematically, we depict the workflow complexity over the setup time required to reach a robust implementation at the desired complexity level. The significant refactoring and restructuring are prone to errors and, therefore, require extensive testing and validation at every scaling stage. By scaling, we refer to increasing the complexity of the three main components: the transition from small to extensive simulations, from simple to complex algorithms, and from laptops to HPC distributed computing systems. 

\FloatBarrier
\QUEENS\ accelerates this process as it already implements mature abstraction layers to allow for scaling to higher levels of complexity. Researchers can, therefore, focus on algorithmic implementations and computational simulations and need to worry less about simulation handling, data management, and interfaces. 

{We summarize our intentions behind \QUEENS\ in the following \emph{mission statement}:

\begin{mission}
Our mission is to provide a comprehensive computational platform for the design and execution of multi-faceted simulation analysis tasks involving data-driven to physics-based computational models. We aim for a high-level abstraction to facilitate rapid prototyping. The platform supports diverse methods, from convergence studies to advanced probabilistic analysis. \QUEENS\ enables incremental algorithm development based on basic Python models to workflows involving sophisticated large-scale (\eg finite element) simulations on HPC platforms. Robust implementations and automated simulation execution should free researchers to focus on their research objectives. We strive for sustainability through the open-source community.
\end{mission}}

The remainder of the paper is structured as follows: We first clarify the target audience and application domains for \QUEENS\ and  provide a short overview of already existing algorithms in Section \ref{sec:overview_analysis}. Afterwards, we outline the architecture and design principles behind the \QUEENS\ software framework in Section \ref{sec:software_design}. We also show how the modular design facilitates the sequential build-up of complex analysis pipelines. Eventually, in the hands-on Section \ref{sec: hands_on},  we demonstrate with the help of two already published examples how a \QUEENS\ simulation can be constructed from scratch. The examples cover a parameter study for a biomechanical tumor growth model and the Bayesian calibration of a fluid-structure interaction problem. Section \ref{sec:conclusion} provides a short conclusion and outlook for the software project \QUEENS.

\FloatBarrier

\section{Overview of currently available methods and target users}
\label{sec:overview_analysis}

Before delving into more detailed architectural aspects, we briefly delineate application domains and the target audience for \QUEENS.
The software is designed for both applied users as well as computational researchers who require an automated framework for analyzing existing simulation frameworks. Typical applications range from simple convergence or automated parameter grid studies to complex optimization tasks or nested probabilistic models and Bayesian inverse analysis in combination with several finite element codes and surrogate models. \QUEENS\ is ideal for users who want to build modular analysis tasks for their existing simulation frameworks (for example, a physics-based finite element model) while having the possibility to \emph{scale} their framework quickly from laptop to HPC clusters. 
The software simplifies the integration of analysis methods with existing legacy codes and manages resources without the user needing to establish these systems from scratch. A distinctive strength of \QUEENS\ is its integrated support for advanced statistical and probabilistic analysis. We enable users to quickly construct probabilistic models and probabilistic machine learning tasks in combination with large numerical simulation frameworks, run uncertainty quantification (UQ) and Bayesian inverse analysis, and integrate surrogate models. In this context, \QUEENS\ is also a suitable framework for \emph{digital twins} in engineering, industry, and science. By \emph{digital twin}, we mean a virtual representation of a (physical) system (\ie a physics-based simulation model discretized by the finite element method) with a bidirectional interaction (\ie repeated flow of data) between virtual representation and real world). 
In such a context, a Bayesian calibration process conducted with \QUEENS can then, for example, update uncertain or unknown parameterizations of the simulation model to reflect the real-world process while accounting for the present uncertainties. The calibrated model can be used for predictive purposes in a subsequent step and can lead to changes in the real system.
\QUEENS\ focuses on efficient parallelization of all mathematical operations and data management.

We currently already offer many state-of-the-art algorithms, ranging from surrogate modeling techniques to UQ to optimization and inference algorithms. We also provide our developments in probabilistic machine learning, UQ, and Bayesian inverse analysis for large-scale simulations. Table \ref{tab:methods} summarizes a collection of currently available methods. As \QUEENS\ is an actively developed code, an up-to-date state of the software can be found in our repository at \repo\ under LGPL license.
\begin{center}
\begin{longtable}{ll}
    \caption{Overview of available methods in \QUEENS.} 
    \label{tab:methods}\\
        \hline
        \rowcolor{lightgray}Algorithm/method & References \\
        \hline
        \endfirsthead
        \multicolumn{2}{c}%
{{\bfseries \tablename\ \thetable{} -- continued from previous page}} \\
\hline
\rowcolor{lightgray}Algorithm/method & References \\ \hline 
\endhead
        \multicolumn{2}{@{}l}{\textbf{Design of experiments}}\\
        Grid designs &  \\
        Convergence studies & \\
        Monte Carlo sampling & \cite{kroese_handbook_2011} \\
        Latin hypercube sampling & \cite{McKay1979} \\
        Sobol sequence & \cite{Sobol1967} \\
        \hline
        \multicolumn{2}{@{}l}{\textbf{Optimization}}\\
        Levenberg-Marquardt algorithm & \cite{Levenberg1944, Marquardt1963} \\
        SciPy optimizers (Conjugate gradient, BFGS, ...) &  \cite{SciPy2020} \\
        Stochastic optimizers (Adam, Adamax, RMSProp) & \cite{Kingma2017, tieleman2012lecture} \\
        \hline
        \multicolumn{2}{@{}l}{\textbf{Uncertainty Quantification}}\\
        Monte Carlo sampling & \cite{kroese_handbook_2011} \\
        Bayesian multi-fidelity Monte Carlo (BMFMC) & \cite{bmfmc_nitzler, multi_fidelity_review, biehler2015towards,koutsourelakis2009accurate} \\
        Polynomial chaos expansion (PCE) & \cite{feinberg_chaospy_2015} \\
        \hline
        \multicolumn{2}{@{}l}{\textbf{Sensitivity analysis}}\\
        Elementary effects method & \cite{Morris1991} \\
        Sobol method & \cite{Sobol1993,Sobol2001} \\
        Sobol method with surrogate uncertainties & \cite{LeGratiet2014,Wirthl2023} \\
        \hline
        \multicolumn{2}{@{}l}{\textbf{Surrogates}} \\
        Gaussian processes (variational, heteroskedastic, etc.) & \cite{Rasmussen2006, Hensman2013} \\
        (Bayesian) Neural networks & \cite{mackay_bayesian_1995}\\
        Gaussian neural networks / inference networks & \\
        Polynomial chaos surrogates & \cite{feinberg_chaospy_2015} \\
        \hline
        \multicolumn{2}{@{}l}{\textbf{(Bayesian) Inverse Analysis (BIA)}}\\
        Markov Chain Monte Carlo & \cite{kroese_handbook_2011, homan_no-u-turn_nodate} \\
        Sequential Monte Carlo & \cite{del_moral_sequential_2006, chopin_introduction_2020}  \\
        Stochastic variational inference (reparameterization trick) & \cite{mohamed_mc_gradient, blei_vi, hoffman_vi, kingma_variational_2015, roeder_vi} \\
        Black box variational inference & \cite{ranganath2014black, arenz_vi, mohamed_mc_gradient}\\
        Bayesian multi-fidelity inverse analysis &\\
        Adaptive and constrained GPs for BIA & \cite{Dinkel_2024}\\
        \hline
\end{longtable}
\end{center}

The methods and algorithms are designed with well-defined interfaces, which makes it very easy to change, combine, or nest different analysis types to build up more complex and flexible algorithms. We will go into more detail on the software architecture's modularity, scalability, and simplicity in the subsequent Section \ref{sec:software_design}.  

\paragraph{First steps and simpler types of analyses (Convergence and grid studies, surrogates)}
A starting point for new \QUEENS\ users might be a small convergence study that evaluates, \eg a finite element model for different FE mesh sizes or polynomial degrees, and afterwards plots an error estimate of the quantity of interest. Other simpler workflows include a two- or three-dimensional parameter study over a parameter grid or specified design of experiments, such as a Latin hypercube design. Section \ref{sec: hands_on} gives a hands-on example of such an analysis. \QUEENS\ evaluates the underlying FE model automatically and in parallel for all combinations of given mesh sizes and polynomial degrees.
Subsequently, the user can directly train and visualize a surrogate model (\eg a Gaussian Process model), which can again be used for further analysis.

\paragraph{Intermediate types of analyses (UQ, sampling methods, optimization, gradient calculations)}
More elaborate analyses, such as optimization, uncertainty quantification (UQ), or global sensitivity analysis, can also be conducted. 
For optimization problems, \QUEENS\ integrates the optimization algorithms provided by SciPy \cite{SciPy2020}, such as the conjugate gradient method, BFGS, and the Levenberg--Marquardt algorithm \cite{Levenberg1944, Marquardt1963}, while taking additional care of embarrassingly parallel tasks and the computation of model gradients. Especially for the latter, we offer a sophisticated abstraction level that enables different ways of computing model gradients via finite difference schemes, automated differentiation (if supported by the simulation model), or incorporating adjoint models to exploit adjoint-based gradient calculation. 
Additionally, \QUEENS\ supports stochastic optimization techniques, such as Adam \cite{Kingma2017}, Adamax \cite{Kingma2017}, and root mean square propagation (RMSProp) \cite{tieleman2012lecture}. 
For UQ, \QUEENS\ provides different approaches to quantify the uncertainty of model outputs: 
besides simple but potentially expensive Monte Carlo methods \cite{kroese_handbook_2011}, \QUEENS\ also supports polynomial chaos expansion (PCE) \cite{feinberg_chaospy_2015} to reduce the number of model evaluations.
If a lower fidelity model is available, \eg a model with fewer degrees of freedom or simplified physics, we also provide our own efficient Bayesian multi-fidelity Monte Carlo method (BMFMC) \cite{bmfmc_nitzler, multi_fidelity_review, biehler2015towards,koutsourelakis2009accurate} to reduce computational cost drastically.
Another intermediate analysis type is the global sensitivity analysis. \QUEENS\ includes global approaches, namely the Elementary Effects method (also called Morris method \cite{Morris1991}) and the Sobol method \cite{Sobol1993, Sobol2001}, to apportion the uncertainty in the model output to different sources of uncertainty in the model input \cite{Saltelli2004}.
Due to the modular design, most methods in \QUEENS\ can use surrogate models to approximate the investigated model, reducing the number of potentially expensive model calls. 
Available regression models range from heteroskedastic Gaussian processes to Bayesian neural networks. The modularity of \QUEENS\ makes it easy to change the analysis type while investigating physics-based simulations, such as finite element models of engineering problems. 

\paragraph{More complex types of analyses (Bayesian inverse analysis and nested probabilistic models)}
More complex algorithmic setups include methods for Bayesian inverse problems in combination with distributed, large-scale physics-based (finite element) models. \QUEENS\ provides a wide range of state-of-the-art inference methods such as Markov Chain Monte Carlo \cite{kroese_handbook_2011} and Sequential Monte Carlo \cite{del_moral_sequential_2006, chopin_introduction_2020} or gradient-based stochastic variational inference \cite{mohamed_mc_gradient, blei_vi, hoffman_vi, kingma_variational_2015, roeder_vi} for which the \QUEENS\ internal gradient abstraction can be used to calculate the model gradients.

\FloatBarrier
\paragraph{Own research in forward and backward uncertainty propagation}
\label{sec:own_research}
In addition to state-of-the-art methods, \QUEENS\ allows us to share our research in Bayesian inverse analysis, uncertainty quantification, optimization, probabilistic machine learning, and global sensitivity analysis tailored to large-scale simulation models. This approach enables other researchers to quickly implement and build upon our methods, fostering collaboration and accelerating progress in the field. The interested reader is encouraged to look into the following publications that present a more in-depth use of \QUEENS\ for complex applications outside this paper's scope.
We, \eg provide novel and very efficient variants of black-box variational inference algorithms  \cite{ranganath2014black, arenz_vi, mohamed_mc_gradient} that do not require model gradients or novel adaptively learned probabilistic surrogates \cite{Dinkel_2024} to approximate the posterior distribution directly and very efficiently. 
For efficient uncertainty quantification, we developed Bayesian multi-fidelity Monte Carlo algorithms \cite{bmfmc_nitzler} based on the authors' previous work \cite{koutsourelakis2009accurate,multi_fidelity_review, biehler2015towards}.
\QUEENS\ also offers methods for global sensitivity analysis in combination with challenging, large-scale biomechanical models \cite{sa_brandstaeter, Wirthl2023}. 
Some other applications of \QUEENS\ include calibration problems involving microstructure development in selective laser melting \cite{microstructure_nitzler, am_meier}, calibration of constitutive laws for biofilms \cite{willmann_levenberg_marquardt}, and tumor-model calibration \cite{kremheller_calibration}, using the Levenberg--Marquardt algorithm. Please also see our publication on Bayesian calibration under uncertainty, when only deformed interface information is accessible \cite{willmann_inverse}, or the Bayesian calibration of an advanced tumor model with actual experimental data \cite{Hervas-Raluy2023}. 

For an up-to-date overview of our publications with \QUEENS\ and further news, please visit our website \url{https://www.queens-py.org}.

\FloatBarrier
\section{Software architecture and design principles}
\label{sec:software_design}
\begin{figure}[htbp]
    \centering
    \includegraphics[width=\textwidth]{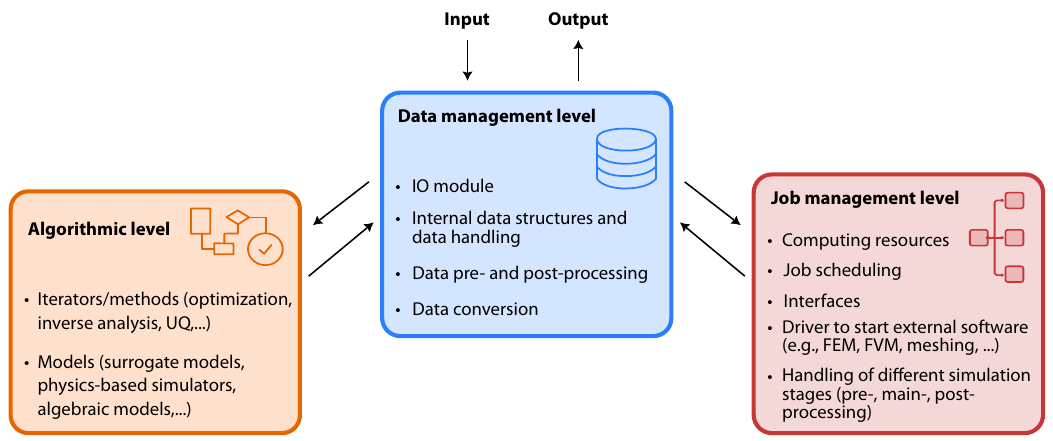}
    \caption{Overview of the \QUEENS\ software architecture and modular design,  reflected in the three principal design levels \emph{algorithmic level}, \emph{data management level}, and the \emph{job management level}.
    }
    \label{fig:architecture}
\end{figure}

To allow for a convenient setup of the algorithms above in combination with a wide range of  (physics-based) simulation codes on different computing infrastructures, \QUEENS\ follows four design principles: \emph{Modularity}, \emph{Scalability}, \emph{Extensibility}, and \emph{Simplicity}. In the following, we give a summary of these principal ideas:

\paragraph{Modularity (flexible and straightforward build-up of workflows from simple blocks)}
We identified an appropriate module abstraction level that balances minimal implementation effort and maximum flexibility. 
The latter allows for easy substitution of components, thereby enhancing the adaptability of the overall system. 
Moreover, it enables the creation of more elaborate nested analysis workflows and encourages using various solvers, software, hardware, and methods to meet specific requirements. 
On the highest abstraction level, we distinguish the \emph{algorithmic level}, the \emph{job management level}, and the \emph{data management level}, which represent a collection of several Python modules in the software architecture. 
Figure~\ref{fig:architecture} overviews these three principal design levels, including a few exemplary sub-modules and their dependencies. 

From the user perspective, most implementations and analyses are carried out at the algorithmic level using \emph{iterators} and \emph{models}, which are the building blocks for creating complex algorithms involving multiple methods and models.
The job management level encompasses the entire process of computation management. Under the term \emph{job}, we understand one pipeline of \emph{setup, pre-processing, scheduling, submitting, post-processing, and data gathering for one physics-based simulation workflow, given one new (parameter) input}. In a broader context, a \emph{job} might also refer to the setup, scheduling, and training of a machine learning model on, \eg GPU resources. This sequence includes interfacing with the required computational resources, scheduling the execution of the different simulation codes (\eg FE simulations or the training of machine learning models), distributing computational tasks, and keeping track of them. Furthermore,  the job management level provides appropriate drivers that let \QUEENS\ communicate with external executables and programs. Additionally, we handle all necessary communication and networking tasks, implement many error detection mechanisms, and log all intermediate steps. This modularity allows us to handle arbitrary simulation codes and execute them via a Dask \cite{dask} backend and their associated input files.

Finally, the data management level is responsible for archiving, storing, and making simulation data available. This level also manages data formats and internal variable structures, handles I/O tasks, ensures data is available across multiple platforms and computing resources, and handles metadata logging.

\begin{figure}[tbp]
    \centering
    \includegraphics[width=\textwidth]{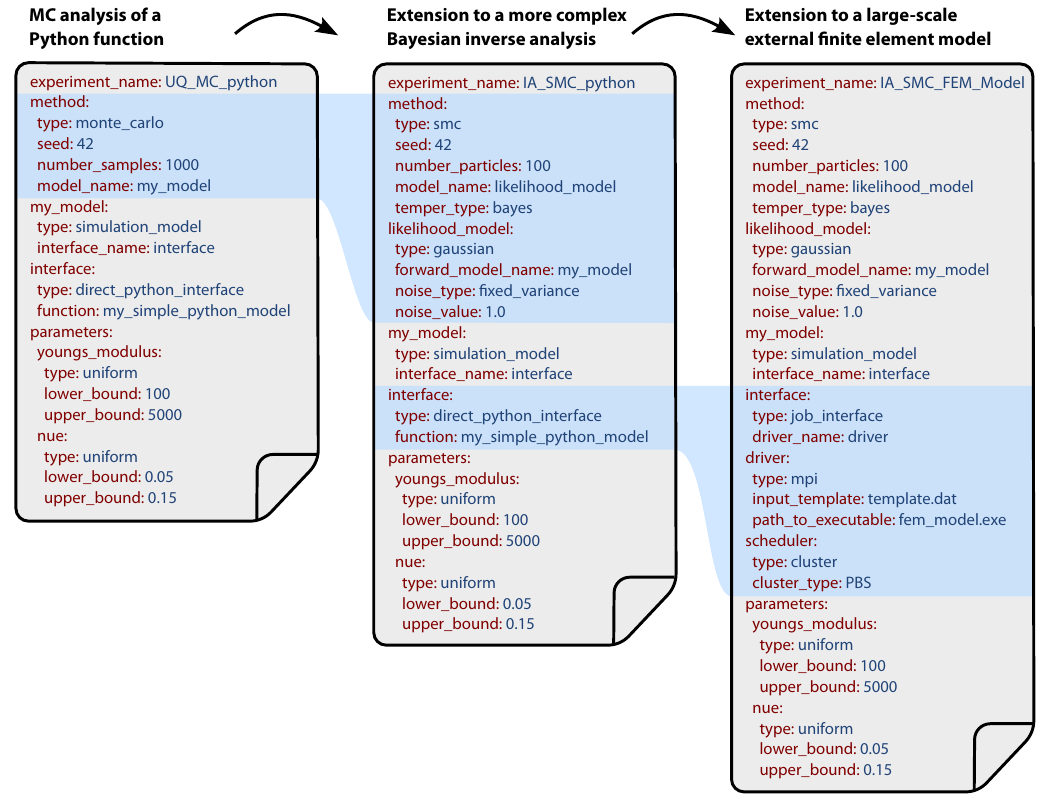}
     \caption{Example of the modular character of a \QUEENS\ input file, which allows for easy extension towards more complex algorithmic setups and types of analysis. We start from a simpler Monte Carlo (MC) analysis of a function written in Python to assess the uncertainty propagation associated with the input parameters \emph{youngs\_modulus} and \emph{nue}. Afterwards, we extend the method or analysis type to a more complex Bayesian inverse analysis. The blue parts of the input file mark the sections that need to be changed. Finally, we replace the Python function with an actual finite element model in the last input file. Please note that we can use any available solver by providing the appropriate interface and the path to the executable.}
    \label{fig:input_files}
\end{figure}

Figure~\ref{fig:input_files} shows how a Monte Carlo analysis of a Python function can be easily changed into a complex Bayesian inverse analysis of an actual large-scale external finite element model by extending and switching building blocks of the analysis. For clarity, we briefly comment on some of the \QUEENS\ internal terms: \emph{Method} denotes the (iterative) algorithm acting on and evaluating an abstract model. The latter can be a \emph{simulation model} (\eg a finite element model) or can be a probabilistic model (\eg a likelihood model), which has another finite element model nested. In principle, a \emph{model} refers to an abstract mapping in a mathematical sense. Another term in the input file is an \emph{interface}. This module specifies how \QUEENS\ can communicate with a \emph{model}: In principle, we distinguish Python internal models, which can be directly accessed and evaluated within the Python programming language, or external models in the form of third-party executables and codes (such as Ansys, other C++ executables, containers, etc.). Finally, the term \emph{driver} means, in this context, a finer-grained version of the interface module that contains settings for the type of MPI communication (local or remote) and handles the parsing of parameters to the respective input files of external executables and codes.

Analysts can quickly and independently change or reuse input blocks from previous \QUEENS\ runs. All building blocks are organized logically and systematically.

\begin{figure}[tbp]
    \centering
    \includegraphics[width=\textwidth]{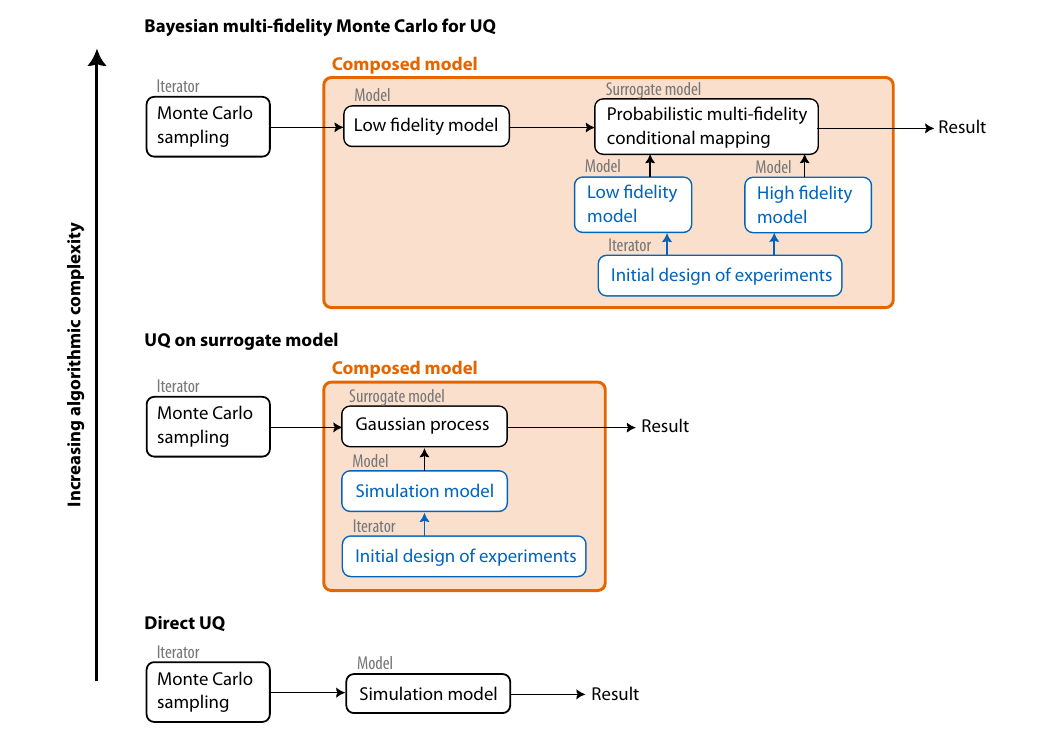}
    \caption{Construction of nested workflows in \QUEENS\ using the example of uncertainty quantification (UQ). 
    The parts marked in blue can be performed offline before starting the UQ.
    In the first case (bottom), we perform direct Monte Carlo sampling on the simulation model (\eg a finite element model) and yield the corresponding output distribution in sample representation. 
    In the second case (middle), we replace the simulation model with a surrogate (\eg a Gaussian process), on which we can then perform the sampling. 
    We initialize and train the surrogate model using an initial design iterator on the more expensive simulation model (\eg a finite element model). 
    Finally, in the top row, we demonstrate the setup of Bayesian multi-fidelity Monte Carlo for UQ: 
    We first conduct Monte Carlo sampling on a low-fidelity model and then correct its response statistically by a probabilistic multi-fidelity mapping. 
    The latter is trained on corresponding tuples of high- and low-fidelity model evaluations drawn from an initial design. 
    }
    \label{fig:nested_workflows}
\end{figure}

\paragraph{Scalability (from simple algorithms on the personal laptop to complex workflows on the HPC cluster)}
We developed \QUEENS\ to manage extensive workloads and intricate algorithms. 
A critical feature enabling this capability is \emph{scalability}. In \QUEENS, scalability appears in multiple forms, such as computational scaling from a laptop to a high-performance computing (HPC) system and algorithmic scaling from a basic parameter study to a comprehensive probabilistic analysis.
Our modular design (see previous paragraph) naturally contributes to scalability. The following paragraph emphasizes key concepts and practical aspects of scalability in \QUEENS.
We intend to nurture a natural incremental workflow in sciences, where problems can be tackled step-wise, from small to large scales, and from simple to complex problems. 
The incremental approach is essential in handling high complexity and growing scientific and technological challenges. 
We utilize the well-established Python library \emph{Dask} \cite{dask} for scheduling several jobs on a local machine or a remote HPC cluster with support for typical queuing systems such as PBS \cite{feng2007-PBSUnifiedPrioritybased} or Slurm \cite{yoo2003-SLURMSimpleLinux}. We also handle the fully distributed parallel execution of individual simulations over multiple processor cores or nodes using a message-passing interface (MPI).
Scaling the computation from a single machine to an HPC cluster is easy because \QUEENS\ automatically handles the tedious steps in the background: Remote communication to the HPC cluster, creating the individual input files and job scripts, submitting jobs to the job queuing system, and gathering the results. We implemented customized versions of many algorithms to ensure embarrassingly parallel execution where possible. Additionally, we included extensive logging capabilities to track the progress of the analysis and simplify debugging.
We achieve the scalability of the algorithmic complexity in \QUEENS\ by utilizing the building blocks of the algorithmic level to switch and especially to nest methods and models. 
These configurations can be easily updated in the input file, allowing for a seamless transition from simple to more complex computations with minimal modifications to the input file.
An example of scaling the algorithmic complexity of uncertainty quantification (UQ) by nesting algorithmic building blocks is presented in Figure~\ref{fig:nested_workflows}. 
    
Many methods and model concepts in \QUEENS\ can exploit parallelization through embarrassingly parallel algorithms or fully distributed simulation frameworks (via the message-passing interface MPI).
    
\paragraph{Extensibility (new methods, collaborations, shared research, and open-source documented scientific code)}
We want the software platform to accommodate future changes and additions, such as adding new functions or modules, without requiring extensive modification of its existing code base. 
In the context of Python, the extensive libraries and modules available in its ecosystem and its simplicity and readability make it easy for developers to add new submodules and functions, making the software more extensible. 
Extensibility for data-driven, probabilistic machine learning and optimization algorithms with many sophisticated and specialized packages is available and under continuous development by the Python community and can be easily included in \QUEENS.

Furthermore, the \QUEENS\ developers are committed to continuously extending and enhancing the software. 
We see \QUEENS\ as a scientific hub where our research is shared, and we encourage other researchers to contribute their methods. 
In this context, \QUEENS\ provides a collaborative scientific platform where researchers can make their methods openly available in the public repository at \repo.
This helps researchers comply with data sharing policies, for example, when publishing papers, accelerates the pace of scientific discovery, and facilitates greater transparency and reproducibility in research.
    
\paragraph{Simplicity (non-expert user to professional method developers)}
We designed \QUEENS\ to be easy to use and to promote understanding through clean syntax, precise and extensive documentation, modular design with well-defined interfaces, and availability of pre-built modules and libraries. 
These abstraction levels make it easy to set up complex workflows and allow straightforward extension and customization with new methods and modules, fostering accessibility and maintainability.

\FloatBarrier

\section{Hands-on examples}
\label{sec: hands_on}

We now give some practical insights on using \QUEENS:
We present the setup of a deterministic parameter study of a tumor-growth model and a Bayesian calibration of a biofilm model, published in \cite{Hervas-Raluy2023, willmann_inverse}.
The algorithmic building blocks can be defined in the \QUEENS\ input file in arbitrary order and then extended as needed. 
The internal setup takes care of the dependencies within the algorithm. 

\subsection{Parameter study of a tumor growth problem with \QUEENS\ }
\label{sec: ParameterStudyTumor}

\begin{figure}[tbp]
    \centering
    \includegraphics[width=\textwidth]{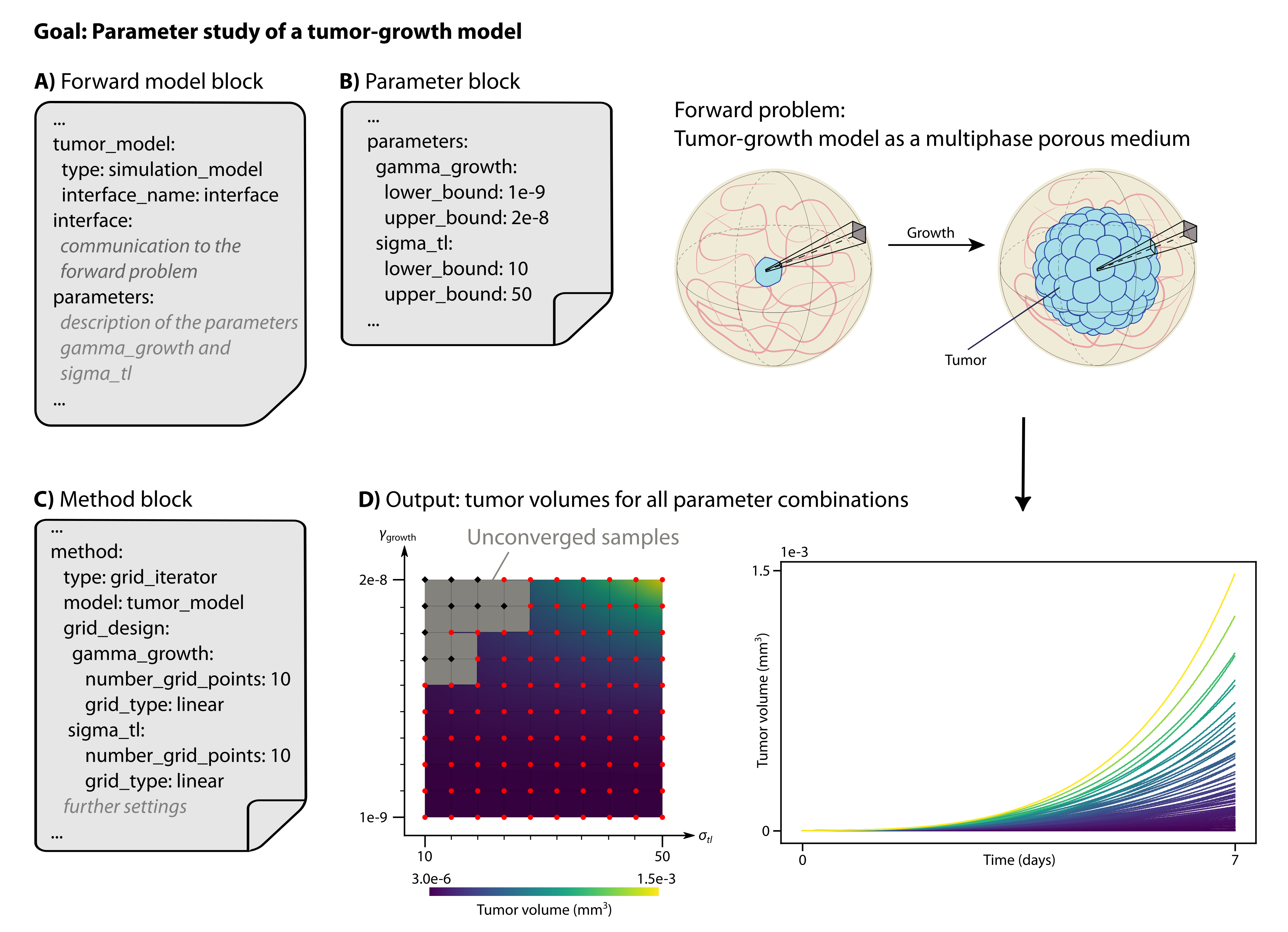}
     \caption{\QUEENS\ setup for a parameter study of a tumor-growth problem.
     The figures of the tumor-growth model are adapted from \cite{Hervas-Raluy2023}.  The red dots in subfigure D) represent converged simulations.}
    \label{fig:ExampleParameterStudyTumor}
\end{figure}

In this example, our goal is a parameter study of a tumor-growth model:
Figure~\ref{fig:ExampleParameterStudyTumor} visualizes the \QUEENS\ setup for this example.

We start by setting up our forward problem.
In this case, we have a tumor growth modeled as a multiphase porous medium and solved with the finite element method (FEM) as, for example, presented in \cite{Hervas-Raluy2023}.
We want to investigate the tumor volume after seven days of growth, depending on a variation of two input parameters: 
the growth coefficient $\gamma_{\textsf{growth}}$ and the surface tension $\sigma_{tl}$. 
In the \QUEENS\ input file, we define the tumor-growth FEM model in the \textbf{forward model block} as shown in Figure~\ref{fig:ExampleParameterStudyTumor}A.
This block configures the type of model (here, a simulation model), the associated parameters (which we will subsequently define in a separate block), and the interface to the simulation software. 
The latter holds paths to the executables and input files for the FEM simulation and other settings, such as MPI and parallelization configurations.

In the \textbf{parameter block} (detailed in Figure~\ref{fig:ExampleParameterStudyTumor}B), we configure the input parameters.
We define the ranges that we want to study for the two parameters.
\QUEENS\ uses the parameter names from the input file and searches for placeholders with the same name in the template input file of the FEM simulation.
It then parses the actual values for these parameters during the \QUEENS\ run to start a new FEM simulation for each parameter combination.

Finally, the \textbf{method block} (detailed in Figure~\ref{fig:ExampleParameterStudyTumor}C) defines which method with which settings to use.
For our parameter study, we use the grid iterator and specify that we want to evaluate a $10 \times 10$ grid for the two input parameters.

The input file has two more building blocks (which are not shown in Figure~\ref{fig:ExampleParameterStudyTumor}): 
A block for global settings, where we set the high-level configurations of the \QUEENS\ run, such as directory paths and temporary directories, and a block for the data management and job scheduling in analogy to Figure~\ref{fig:input_files}.

After the setup, we start the \QUEENS\ run and let the software configure and build up the workflow. 
A live update of which analysis section is currently running is displayed on the terminal output.
The analysis results are saved as a pickle file containing the evaluated input parameter combinations with the resulting tumor volumes.
If desired, \QUEENS\ can plot and visualize parts of the analysis depending on the settings and flags set in the input file. 

In our example, the output (Figure~\ref{fig:ExampleParameterStudyTumor}D) shows that the largest tumor volume is associated with the largest value for both the growth coefficient and the surface tension.
We additionally find that for several parameter combinations, the FEM simulation did not converge:
Convergence is not reached for small values of the surface tension combined with large values of the growth coefficient, marked in gray in Figure~\ref{fig:ExampleParameterStudyTumor}.
Hence, our parameter study directly included parts of a convergence study.
We can also visualize the tumor volume over time for all parameter combinations to gain more insight into the results of our analysis.

\subsection{Bayesian calibration of a fluid-structure interaction problem with \QUEENS\ }
\label{sec: SmcBiofilm}

\begin{figure}[tbp]
    \centering
    \includegraphics[width=\textwidth]{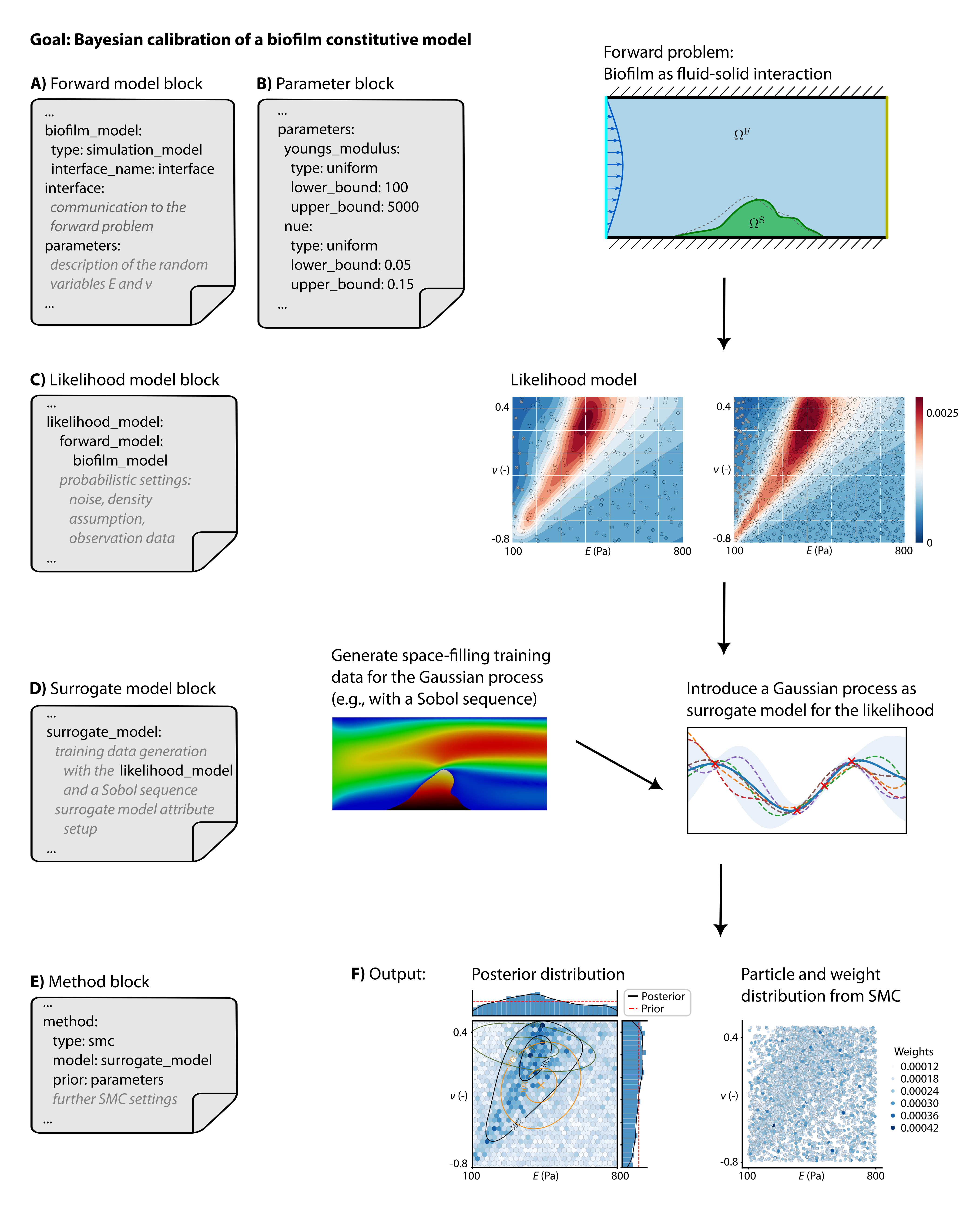}
     \caption{\QUEENS\ setup for a Bayesian calibration of a coupled fluid-structure interaction problem (FEM simulation) using a Gaussian process meta-model to mitigate the computational costs.
     Figures adapted from \cite{willmann_inverse}.}
    \label{fig:ExampleBiofilmIA}
\end{figure}

We now discuss the setup of a Bayesian calibration problem for a coupled biofilm finite element simulation based on a recent publication of the authors \cite{willmann_inverse}. 
Figure~\ref{fig:ExampleBiofilmIA} presents an overview of the input file setup and the workflow.

We can formulate the Bayesian calibration problem as follows: 
Given is a pipe or channel through which a fluid flows with an (uncertain) velocity profile. 
A biofilm grows on the lower boundary of the channel. 
A camera installation lets us observe the biofilm's boundary or interface deformation due to the surrounding fluid flow. 
This kind of scenario has applications in many scientific and industrial settings. 
For further analysis, we are interested in the constitutive properties of the biofilm, \eg to make predictions when pipes in chemical industry plants clog due to biofilm growth.
We have a coupled fluid-solid interaction finite element problem that virtually describes the scenario. 
The solid domain has a constitutive model with unknown Young's modulus $E$ and Poisson's ratio $\nu$. 
We want to infer these parameters based on the experimentally measured interface deformations. 
Figure~\ref{fig:ExampleBiofilmIA} overviews the problem and the corresponding Bayesian inverse problem setup in \QUEENS. 
The figure guides the analysis setup by input file snippets that contain pseudo code and explanatory plots corresponding to these building blocks. 
It is easiest to start from the computational forward model, which is what we do in this example: 
As shown in Figure~\ref{fig:ExampleBiofilmIA}A, we define the biofilm model in the \textbf{forward model block}, similar to the previously discussed tumor-growth model example.
The \textbf{parameter block} defines the uncertain input parameters (Young's modulus $E$ and Poisson's ratio $\nu$) and their prior distributions (see Figure~\ref{fig:ExampleBiofilmIA}B).
We here use uniform prior distributions.

In the next step, we define the likelihood model for the inverse analysis in the \textbf{likelihood model block}, as detailed in Figure~\ref{fig:ExampleBiofilmIA}C.
\QUEENS\ offers several types of likelihood models. 
In this example, we choose a standard Gaussian likelihood with fixed and static noise variance. 
\QUEENS\ offers more options, such as adaptive likelihood noise with iterative averaging, which we omit here for simplicity. 
The likelihood model block furthermore requires a path to an external file (\eg a CSV file) that holds the experimental measurements along with their coordinates as observation data. 
We also specify which forward model the likelihood model is based upon. 
In our case, it is the previously defined biofilm FEM model. 
Together with the parameter block, which provides the prior model, this defines the full probabilistic model for the Bayesian inverse analysis, according to Bayes' rule.

As evaluating the likelihood model triggers an expensive forward simulation of the coupled FEM biofilm model, we choose to replace the mapping of the likelihood model with a cheaper-to-evaluate surrogate model. 
\QUEENS\ offers a wide range of possible surrogate models. 
Here, we select a Gaussian process model, which we set up in a separate \textbf{surrogate model block} (see Figure~\ref{fig:ExampleBiofilmIA}D). 
Within this block, we provide an iterator that specifies how (with which method and from which model) and how many training data points shall be generated to train it. 
We can, \eg select a space-filling iterator, such as the Sobol sequence, set the number of training points to 500, and specify the likelihood model as the target for the training points. 
Once we have started the \QUEENS\ run, the software allocates the resources we specified for the FEM model (the building block of the biofilm model) and runs simulations in parallel until it has realized the initial training data. 
Afterwards, the surrogate is trained according to other settings, such as the selected optimizer, chosen kernel model, and mean function. 
During the analysis, \QUEENS\ can also provide metrics that assess the surrogate's quality to refine the latter if necessary.
Note that in \QUEENS, simulations models, surrogate models, and likelihood models are all models a method can act on.

Eventually, we can finish the algorithmic setup by defining the \textbf{method block} as shown in Figure~\ref{fig:ExampleBiofilmIA}E.
Here, we choose a sequential Monte Carlo method (SMC) to solve the Bayesian inverse problem. 
We set the Gaussian process meta-model as the model the iterator should act upon, and also hand over the parameter block for the prior definition.
We also configure method-specific settings, such as tempering strategy, effective sample size (ESS), number of particles, and MCMC chains.

After the setup, we can start the \QUEENS\ run and let the software configure, build up, and run the complex workflow. 
The final result, in this case, the posterior distribution of the uncertain Young's modulus $E$ and Poisson's ratio $\nu$ will be saved as a pickle file in a sample representation along with valuable statistics and further meta-data and information, partly depending on the individual module setups in the input file. 
Figure~\ref{fig:ExampleBiofilmIA}F presents a possible output visualization.

\FloatBarrier

\section{Conclusion}
\label{sec:conclusion}

We presented \QUEENS\ as a novel comprehensive software platform for coordinating and composing complex analysis workflows for arbitrary simulation software.
\QUEENS\ is an open-source Python framework built on modularity, scalability, extensibility, and simplicity.
Together with the variety of deterministic and probabilistic methods in combination with arbitrary forward models, this encourages the use and further development of \QUEENS\ in the research community and industry.
Thus, researchers circumvent time-consuming pitfalls and challenges when setting up a complex workflow, such as running a complex probabilistic algorithm with a computationally expensive forward model on an HPC cluster.

In the coming years, we will continually update the software with a focus on our research in probabilistic analysis. We will make our published methods in UQ, Bayesian inverse problems, sensitivity analysis, and physics-informed machine learning available through \QUEENS\ while striving for applicability for large-scale numerical problems. We invite other interested groups and researchers to use and actively contribute to \QUEENS\ in our repository, which can be found under \repo, along with complete documentation of the code and continuous integration pipelines.

If you want to use \QUEENS\ in your published research, it is mandatory, given the generally accepted guidelines for good scientific practice, to cite this \QUEENS\ software paper.

\section*{Notes on origin and evolution of \QUEENS}
The software project \QUEENS\ was initiated in 2016 by AdCo $\mathrm{Engineering}^{\mathrm{GW}}$ GmbH (abbreviated as “AdCo” in the following). From 2017 through 2020, with JB being the lead developer at AdCo, it was supported by the Bavarian Ministry of Economic Affairs, Regional Development and Energy via their program BayTOU for promoting technology-oriented start-ups, as will be detailed in the subsequent Acknowledgements. In 2019, a collaboration with the Institute for Computational Mechanics of the Technical University of Munich (TUM) was started, aiming at further development of the software with respect to new innovative methods for uncertainty quantification, optimization, inverse analysis, etc. Later on, on the one hand, \QUEENS\ was made available by AdCo to the TUM Institute for Computational Mechanics for redesign and further expansion (meanwhile as an Open Source code together with the Institute for Mathematics and Computer-Based Simulation of the University of the Bundeswehr Munich). On the other hand, starting from \QUEENS, the commercial software SQUEENS (“Software for Quantifying Uncertain Effects in Engineering Systems”) was developed and established, among others, enabling the use of widely used commercial CAE software packages (such as ANSYS) as well as open-source CAE software packages (such as OpenFOAM) within SQUEENS and featuring a convenient graphical user interface (GUI). On 16 November 2021, SQUEENS was registered as a trademark and has been continuously supported by AdCo to date.

\section*{Acknowledgments}
We gratefully acknowledge the initial funding for \QUEENS, received by \emph{AdCo Engineering$^{\text{GW}}$ GmbH} within the scope of the BayTOU project QUEENS (project no TOU-1606-0003) supported by the \emph{Bavarian Ministry of Economic Affairs, Regional Development and Energy} through its program \emph{BayTOU}. We furthermore want to thank the following funding sources that allowed us to continue the work on \QUEENS: JN acknowledges support from the German Research Foundation (DFG) within the priority program SPP 1886.  Furthermore, WAW, MD, and JN acknowledge support from the European Research Council (grant agreement No. 101021526-BREATHE). GRR acknowledges support from the German Federal Ministry of Education and Research (FestBatt 2 03XP0435B).
SB gratefully acknowledges funding by the Deutsche Forschungsgemeinschaft (DFG, German Research Foundation), project number 257981274.
Finally,  we want to thank our former students and collaborators.

\bibliography{main.bbl}

\begin{thebibliography}{10}
\providecommand \doibase [0]{http://dx.doi.org/}%

\bibitem{pydoe}
Baudin M, Christopoulou M, Collette Y, Martinez JM. py{DOE}: The experimental design package for Python.;  2013.

\bibitem{Iwanaga2022}
Iwanaga T, Usher W, Herman J. Toward {SALib} 2.0: {Advancing} the accessibility and interpretability of global sensitivity analyses. {\it Socio-Environmental Systems Modelling} 2022\string; 4\string: 18155.
\newblock \href {\doibase 10.18174/sesmo.18155} {doi: 10.18174/sesmo.18155}

\bibitem{feinberg_chaospy_2015}
Feinberg J, Langtangen HP. Chaospy: {An} open source tool for designing methods of uncertainty quantification. {\it Journal of Computational Science} 2015\string; 11\string: 46--57.
\newblock \href {\doibase 10.1016/j.jocs.2015.08.008} {doi: 10.1016/j.jocs.2015.08.008}

\bibitem{marelli2014-UQLabFrameworkUncertainty}
Marelli S, Sudret B. {UQLab}: A Framework for Uncertainty Quantification in Matlab. In: The 2nd International Conference on Vulnerability and Risk Analysis and Management (ICVRAM 2014). ; 2014; University of Liverpool, United Kingdom\string: 2554--2563

\bibitem{olivier2020-UQpyGeneralPurpose}
Olivier A, Giovanis DG, Aakash BS, Chauhan M, Vandanapu L, Shields MD. {{UQpy}}: {{A}} General Purpose {{Python}} Package and Development Environment for Uncertainty Quantification. {\it Journal of Computational Science} 2020\string; 47\string: 101204.
\newblock \href {\doibase 10.1016/j.jocs.2020.101204} {doi: 10.1016/j.jocs.2020.101204}

\bibitem{riis2024cuqipy}
Riis NAB, Alghamdi AMA, Uribe F, et al. CUQIpy: I. Computational uncertainty quantification for inverse problems in Python. {\it Inverse Problems} 2024\string; 40(4)\string: 045009.
\newblock \href {\doibase 10.1088/1361-6420/ad22e7} {doi: 10.1088/1361-6420/ad22e7}

\bibitem{alghamdi2024cuqipy}
Alghamdi AMA, Riis NAB, Afkham BM, et al. CUQIpy: II. Computational uncertainty quantification for PDE-based inverse problems in Python. {\it Inverse Problems} 2024\string; 40(4)\string: 045010.
\newblock \href {\doibase 10.1088/1361-6420/ad22e8} {doi: 10.1088/1361-6420/ad22e8}

\bibitem{Lintusaari2018}
Lintusaari J, Vuollekoski H, Kangasr{\"a}{\"a}si{\"o} A, et al. ELFI: Engine for Likelihood-Free Inference. {\it Journal of Machine Learning Research} 2018\string; 19(16)\string: 1-7.

\bibitem{pymc2023}
Oriol AP, Virgile A, Colin C, et al. PyMC: A Modern and Comprehensive Probabilistic Programming Framework in Python. {\it {PeerJ} Computer Science} 2023\string; 9\string: e1516.
\newblock \href {\doibase 10.7717/peerj-cs.1516} {doi: 10.7717/peerj-cs.1516}

\bibitem{chopin_introduction_2020}
Chopin N, Papaspiliopoulos O. {\it An Introduction to Sequential Monte Carlo}.
\newblock Springer International Publishing .
\newblock 2020

\bibitem{seelinger2023-UMBridgeUncertaintyQuantification}
Seelinger L, Cheng-Seelinger V, Davis A, Parno M, Reinarz A. {{UM-Bridge}}: {{Uncertainty}} Quantification and Modeling Bridge. {\it Journal of Open Source Software} 2023\string; 8(83)\string: 4748.
\newblock \href {\doibase 10.21105/joss.04748} {doi: 10.21105/joss.04748}

\bibitem{dakota}
Adams BM, Bohnhoff WJ, Dalbey KR, et al. Dakota, A Multilevel Parallel Object-Oriented Framework for Design Optimization, Parameter Estimation, Uncertainty Quantification, and Sensitivity Analysis (V.6.16 User's Manual). tech. rep., Sandia National Laboratories (SNL-NM); Albuquerque, NM (United States):   2021

\bibitem{baudin2017-OpenTURNSIndustrialSoftware}
Baudin M, Dutfoy A, Iooss B, Popelin AL. {{OpenTURNS}}: {{An Industrial Software}} for {{Uncertainty Quantification}} in {{Simulation}}. In:  Ghanem R, Higdon D, Owhadi H. \kern-2pt, eds. {\it Handbook of {{Uncertainty Quantification}}}{Cham}: {Springer International Publishing}.  2017 (pp. 2001--2038)

\bibitem{Richardson2020_easyvvuq}
Richardson RA, Wright DW, Edeling W, Jancauskas V, Lakhlili J, Coveney PV. EasyVVUQ: A Library for Verification, Validation and Uncertainty Quantification in High Performance Computing. {\it Journal of Open Research Software} 2020.
\newblock \href {\doibase 10.5334/jors.303} {doi: 10.5334/jors.303}

\bibitem{kroese_handbook_2011}
Kroese DP, Taimre T, Botev ZI. {\it Handbook of {Monte} {Carlo} {Methods}}.
\newblock Wiley {Series} in {Probability} and {Statistics}Wiley.
\newblock 1~ed. 2011

\bibitem{McKay1979}
McKay MD, Beckman RJ, Conover WJ. A {{Comparison}} of {{Three Methods}} for {{Selecting Values}} of {{Input Variables}} in the {{Analysis}} of {{Output}} from a {{Computer Code}}. {\it Technometrics} 1979\string; 21(2)\string: 239--245.
\newblock \href {\doibase 10.2307/1268522} {doi: 10.2307/1268522}

\bibitem{Sobol1967}
Sobol' I. On the Distribution of Points in a Cube and the Approximate Evaluation of Integrals. {\it USSR Computational Mathematics and Mathematical Physics} 1967\string; 7(4)\string: 86--112.
\newblock \href {\doibase 10.1016/0041-5553(67)90144-9} {doi: 10.1016/0041-5553(67)90144-9}

\bibitem{Levenberg1944}
Levenberg K. A Method for the Solution of Certain Non-Linear Problems in Least Squares. {\it Quarterly of Applied Mathematics} 1944\string; 2(2)\string: 164--168.
\newblock \href {\doibase 10.1090/qam/10666} {doi: 10.1090/qam/10666}

\bibitem{Marquardt1963}
Marquardt DW. An {{Algorithm}} for {{Least-Squares Estimation}} of {{Nonlinear Parameters}}. {\it Journal of the Society for Industrial and Applied Mathematics} 1963\string; 11(2)\string: 431--441.
\newblock \href {\doibase 10.1137/0111030} {doi: 10.1137/0111030}

\bibitem{SciPy2020}
Virtanen P, Gommers R, Oliphant TE, et al. {{SciPy} 1.0: Fundamental Algorithms for Scientific Computing in Python}. {\it Nature Methods} 2020\string; 17\string: 261--272.
\newblock \href {\doibase 10.1038/s41592-019-0686-2} {doi: 10.1038/s41592-019-0686-2}

\bibitem{Kingma2017}
Kingma DP, Ba J. Adam: A Method for Stochastic Optimization.;  2017.

\bibitem{tieleman2012lecture}
Tieleman T, Hinton G, others . Lecture 6.5-rmsprop: Divide the gradient by a running average of its recent magnitude. {\it COURSERA: Neural networks for machine learning} 2012\string; 4(2)\string: 26--31.

\bibitem{bmfmc_nitzler}
Nitzler J, Biehler J, Fehn N, Koutsourelakis PS, Wall WA. A generalized probabilistic learning approach for multi-fidelity uncertainty quantification in complex physical simulations. {\it Computer Methods in Applied Mechanics and Engineering} 2022\string; 400\string: 115600.
\newblock \href {\doibase https://doi.org/10.1016/j.cma.2022.115600} {doi: https://doi.org/10.1016/j.cma.2022.115600}

\bibitem{multi_fidelity_review}
Biehler J, Mäck M, Nitzler J, Hanss M, Koutsourelakis PS, Wall WA. Multifidelity approaches for uncertainty quantification. {\it GAMM-Mitteilungen} 2019\string; 42(2)\string: e201900008.
\newblock \href {\doibase https://doi.org/10.1002/gamm.201900008} {doi: https://doi.org/10.1002/gamm.201900008}

\bibitem{biehler2015towards}
Biehler J, Gee MW, Wall WA. Towards efficient uncertainty quantification in complex and large-scale biomechanical problems based on a Bayesian multi-fidelity scheme. {\it Biomechanics and modeling in mechanobiology} 2015\string; 14\string: 489--513.
\newblock \href {\doibase 10.1007/s10237-014-0618-0} {doi: 10.1007/s10237-014-0618-0}

\bibitem{koutsourelakis2009accurate}
Koutsourelakis PS. Accurate uncertainty quantification using inaccurate computational models. {\it SIAM Journal on Scientific Computing} 2009\string; 31(5)\string: 3274--3300.
\newblock \href {\doibase 10.1137/080733565} {doi: 10.1137/080733565}

\bibitem{Morris1991}
Morris MD. Factorial {{Sampling Plans}} for {{Preliminary Computational Experiments}}. {\it Technometrics} 1991\string; 33(2)\string: 161--174.
\newblock \href {\doibase 10.1080/00401706.1991.10484804} {doi: 10.1080/00401706.1991.10484804}

\bibitem{Sobol1993}
Sobol IM. Sensitivity {{Estimates}} for {{Nonlinear Mathematical Models}}. {\it Mathematical modelling and computational experiments} 1993\string; 1(4)\string: 407--414.

\bibitem{Sobol2001}
Sobol IM. Global Sensitivity Indices for Nonlinear Mathematical Models and Their {{Monte Carlo}} Estimates. {\it Mathematics and Computers in Simulation} 2001\string; 55(1)\string: 271--280.
\newblock \href {\doibase 10.1016/S0378-4754(00)00270-6} {doi: 10.1016/S0378-4754(00)00270-6}

\bibitem{LeGratiet2014}
Le~Gratiet L, Cannamela C, Iooss B. A {{Bayesian Approach}} for {{Global Sensitivity Analysis}} of ({{Multifidelity}}) {{Computer Codes}}. {\it SIAM/ASA Journal on Uncertainty Quantification} 2014\string; 2(1)\string: 336--363.
\newblock \href {\doibase 10.1137/130926869} {doi: 10.1137/130926869}

\bibitem{Wirthl2023}
Wirthl B, Brandstaeter S, Nitzler J, Schrefler BA, Wall WA. Global Sensitivity Analysis Based on {{Gaussian-process}} Metamodelling for Complex Biomechanical Problems. {\it International Journal for Numerical Methods in Biomedical Engineering} 2023\string; 39(3)\string: e3675.
\newblock \href {\doibase 10.1002/cnm.3675} {doi: 10.1002/cnm.3675}

\bibitem{Rasmussen2006}
Rasmussen CE, Williams CKI. {\it Gaussian {{Processes}} for {{Machine Learning}}}.
\newblock {Cambridge, Massachusetts, London, England}: {The MIT Press} .
\newblock 2006.

\bibitem{Hensman2013}
Hensman J, Fusi N, Lawrence ND. Gaussian Processes for {{Big}} Data. {\it Proceedings of the {{Twenty-Ninth Conference}} on {{Uncertainty}} in {{Artificial Intelligence}} (UAI2013)} 2013\string; 1309.6835\string: 282--290.

\bibitem{mackay_bayesian_1995}
MacKay DJ. Bayesian neural networks and density networks. {\it Nuclear Instruments and Methods in Physics Research Section A: Accelerators, Spectrometers, Detectors and Associated Equipment} 1995\string; 354(1)\string: 73--80.
\newblock \href {\doibase 10.1016/0168-9002(94)00931-7} {doi: 10.1016/0168-9002(94)00931-7}

\bibitem{homan_no-u-turn_nodate}
Hoffman MD, Gelman A. The no-u-turn sampler: Adaptively setting path lengths in hamiltonian monte carlo. {\it Journal of Machine Learning Research} 2014\string; 15(47)\string: 1593--1623.

\bibitem{del_moral_sequential_2006}
Del~Moral P, Doucet A, Jasra A. Sequential {Monte} {Carlo} samplers. {\it Journal of the Royal Statistical Society: Series B (Statistical Methodology)} 2006\string; 68(3)\string: 411--436.
\newblock \href {\doibase 10.1111/j.1467-9868.2006.00553.x} {doi: 10.1111/j.1467-9868.2006.00553.x}

\bibitem{mohamed_mc_gradient}
Mohamed S, Rosca M, Figurnov M, Mnih A. Monte Carlo Gradient Estimation in Machine Learning. {\it Journal of Machine Learning Research} 2020\string; 21(132)\string: 1--62.

\bibitem{blei_vi}
Blei DM, Kucukelbir A, McAuliffe JD. Variational Inference: A Review for Statisticians. {\it Journal of the American Statistical Association} 2017\string; 112(518)\string: 859-877.
\newblock \href {\doibase 10.1080/01621459.2017.1285773} {doi: 10.1080/01621459.2017.1285773}

\bibitem{hoffman_vi}
Hoffman MD, Blei DM, Wang C, Paisley J. Stochastic Variational Inference. {\it Journal of Machine Learning Research} 2013\string; 14(40)\string: 1303--1347.

\bibitem{kingma_variational_2015}
Kingma DP, Salimans T, Welling M. Variational {Dropout} and the {Local} {Reparameterization} {Trick}. In:  Cortes C, Lawrence N, Lee D, Sugiyama M, Garnett R. \kern-2pt, eds. {\it Advances in {Neural} {Information} {Processing} {Systems}}. 28. Curran Associates, Inc. ; 2015.

\bibitem{roeder_vi}
Roeder G, Wu Y, Duvenaud DK. Sticking the Landing: Simple, Lower-Variance Gradient Estimators for Variational Inference. In:  Guyon I, Luxburg UV, Bengio S, et al. \kern-2pt, eds. {\it Advances in Neural Information Processing Systems}. 30. Curran Associates, Inc. ; 2017.

\bibitem{ranganath2014black}
Ranganath R, Gerrish S, Blei D. Black box variational inference. In: PMLR. ; 2014\string: 814--822.

\bibitem{arenz_vi}
Arenz O, Neumann G, Zhong M. Efficient Gradient-Free Variational Inference using Policy Search. In:  Dy J, Krause A. \kern-2pt, eds. {\it Proceedings of the 35th International Conference on Machine Learning}. 80 of {\it Proceedings of Machine Learning Research}. PMLR. ; 2018\string: 234--243.

\bibitem{Dinkel_2024}
Dinkel M, Geitner CM, Rei GR, Nitzler J, Wall WA. Solving Bayesian inverse problems with expensive likelihoods using constrained Gaussian processes and active learning. {\it Inverse Problems} 2024\string; 40(9)\string: 095008.
\newblock \href {\doibase 10.1088/1361-6420/ad5eb4} {doi: 10.1088/1361-6420/ad5eb4}

\bibitem{Saltelli2004}
Saltelli A. \kern-2pt, ed.{\it Sensitivity Analysis in Practice: A Guide to Assessing Scientific Models}.
\newblock {Hoboken, NJ}: {Wiley} .
\newblock 2004.

\bibitem{sa_brandstaeter}
Brandstaeter S, Fuchs SL, Biehler J, Aydin RC, Wall WA, Cyron CJ. Global sensitivity analysis of a homogenized constrained mixture model of arterial growth and remodeling. {\it Journal of Elasticity} 2021\string; 145(1-2)\string: 191--221.
\newblock \href {\doibase https://doi.org/10.1007/s10659-021-09833-9} {doi: https://doi.org/10.1007/s10659-021-09833-9}

\bibitem{microstructure_nitzler}
Nitzler J, Meier C, M{\"u}ller KW, Wall WA, Hodge NE. A novel physics-based and data-supported microstructure model for part-scale simulation of laser powder bed fusion of Ti-6Al-4V. {\it Advanced Modeling and Simulation in Engineering Sciences} 2021\string; 8(1)\string: 16.
\newblock \href {\doibase https://doi.org/10.1186/s40323-021-00201-9} {doi: https://doi.org/10.1186/s40323-021-00201-9}

\bibitem{am_meier}
Meier C, Fuchs SL, Much N, et al. Physics-based modeling and predictive simulation of powder bed fusion additive manufacturing across length scales. {\it GAMM-Mitteilungen} 2021\string; 44(3)\string: e202100014.
\newblock \href {\doibase https://doi.org/10.1002/gamm.202100014} {doi: https://doi.org/10.1002/gamm.202100014}

\bibitem{willmann_levenberg_marquardt}
Willmann H, Wall WA. Inverse analysis of material parameters in coupled multi-physics biofilm models. {\it Advanced Modeling and Simulation in Engineering Sciences} 2022\string; 9(1)\string: 1--32.
\newblock \href {\doibase https://doi.org/10.1186/s40323-022-00220-0} {doi: https://doi.org/10.1186/s40323-022-00220-0}

\bibitem{kremheller_calibration}
Kremheller J, Brandstaeter S, Schrefler BA, Wall WA. Validation and parameter optimization of a hybrid embedded/homogenized solid tumor perfusion model. {\it International Journal for Numerical Methods in Biomedical Engineering} 2021\string; 37(8)\string: e3508.
\newblock \href {\doibase https://doi.org/10.1002/cnm.3508} {doi: https://doi.org/10.1002/cnm.3508}

\bibitem{willmann_inverse}
Willmann H, Nitzler J, Brandst{\"a}ter S, Wall WA. Bayesian calibration of coupled computational mechanics models under uncertainty based on interface deformation. {\it Advanced Modeling and Simulation in Engineering Sciences} 2022\string; 9(1)\string: 24.
\newblock \href {\doibase https://doi.org/10.1186/s40323-022-00237-5} {doi: https://doi.org/10.1186/s40323-022-00237-5}

\bibitem{Hervas-Raluy2023}
{Hervas-Raluy} S, Wirthl B, Guerrero PE, et al. Tumour growth: An approach to calibrate parameters of a multiphase porous media model based on in vitro observations of Neuroblastoma spheroid growth in a hydrogel microenvironment. {\it Computers in Biology and Medicine} 2023\string; 159\string: 106895.
\newblock \href {\doibase 10.1016/j.compbiomed.2023.106895} {doi: 10.1016/j.compbiomed.2023.106895}

\bibitem{dask}
Developers DC. Dask: Library for dynamic task scheduling.;  2016.

\bibitem{feng2007-PBSUnifiedPrioritybased}
Feng H, Misra V, Rubenstein D. {{PBS}}: A Unified Priority-Based Scheduler. {\it SIGMETRICS Perform. Eval. Rev.} 2007\string; 35(1)\string: 203--214.
\newblock \href {\doibase 10.1145/1269899.1254906} {doi: 10.1145/1269899.1254906}

\bibitem{yoo2003-SLURMSimpleLinux}
Yoo AB, Jette MA, Grondona M. {{SLURM}}: {{Simple}} Linux Utility for Resource Management. In:  Feitelson D, Rudolph L, Schwiegelshohn U. \kern-2pt, eds. {\it Job Scheduling Strategies for Parallel Processing}Springer Berlin Heidelberg; 2003\string: 44--60.

\end{thebibliography}
\end{document}